\documentclass[osajnl,twocolumn,showpacs,superscriptaddress,10pt]{revtex4-1}
\usepackage{graphicx}
\usepackage{amsmath}
\usepackage{amssymb}
\begin{document}
\title{Counter-rotating cavity solitons in a silicon nitride microresonator}

\author{Chaitanya~Joshi}\email{Corresponding author: chaitanya.joshi@columbia.edu}
\affiliation{Department of Applied Physics and Applied Mathematics, Columbia University, New York, NY 10027, USA}
\affiliation{School of Applied and Engineering Physics, Cornell University, Ithaca, NY 14853, USA}

\author{Alexander~Klenner}
\affiliation{Department of Applied Physics and Applied Mathematics, Columbia University, New York, NY 10027, USA}

\author{Yoshitomo~Okawachi}
\affiliation{Department of Applied Physics and Applied Mathematics, Columbia University, New York, NY 10027, USA}

\author{Mengjie~Yu}
\affiliation{Department of Applied Physics and Applied Mathematics, Columbia University, New York, NY 10027, USA}
\affiliation{School of Electrical and Computer Engineering, Cornell University, Ithaca, NY 14853, USA}

\author{Kevin~Luke}
\affiliation{School of Electrical and Computer Engineering, Cornell University, Ithaca, NY 14853, USA}

\author{Xingchen~Ji}
\affiliation{School of Electrical and Computer Engineering, Cornell University, Ithaca, NY 14853, USA}
\affiliation{Department of Electrical Engineering, Columbia University, New York, NY 10027, USA}

\author{Michal~Lipson}
\affiliation{Department of Electrical Engineering, Columbia University, New York, NY 10027, USA}

\author{Alexander~L.~Gaeta}
\affiliation{Department of Applied Physics and Applied Mathematics, Columbia University, New York, NY 10027, USA}





\begin{abstract}We demonstrate the generation of counter-rotating cavity solitons in a silicon nitride microresonator using a fixed, single-frequency laser. We demonstrate a dual 3-soliton state with a difference in the repetition rates of the soliton trains that can be tuned by varying the ratio of pump powers in the two directions. Such a system enables a highly compact, tunable dual comb source that can be used for applications such as spectroscopy and  distance ranging.
\end{abstract}

\maketitle

\noindent
Advancements in optical frequency comb technology over the past two decades have enabled applications in a wide range of fields including precision spectroscopy \cite{diddams2007}, frequency metrology \cite{udem2002}, optical clockwork \cite{diddams2001,newbury2011}, astronomical spectrograph calibration \cite{li2008,steinmetz2008}, and microwave signal synthesis \cite{fortier2011}. Applications benefit from the high precision of the frequencies of the comb lines and require low noise, stable operation \cite{ye2007}. Stabilized low-noise comb sources were first demonstrated using mode-locked solid state lasers and fiber lasers \cite{diddams2000,jones2000}. Over the last decade, on-chip optical frequency comb generation using microresonators has seen significant progress and has been demonstrated in several materials including silica \cite{delhaye2007,li2012,yi2015,webb2016}, crystalline fluorides \cite{savchenkov2008,liang2011,herr2014}, silicon nitride (Si$_3$N$_4$) \cite{foster2011,wang2013,joshi2016,brasch2016,li2017}, hydex \cite{Peccianti2012}, diamond \cite{hausmann2014}, aluminum nitride \cite{jung2014}, silicon \cite{griffith2015,yu2016}, and AlGaAs \cite{pu2016}. Low-noise soliton mode-locked microresonator frequency combs have been demonstrated \cite{herr2014,wang2016,li2017,brasch2016,yi2015,webb2016,joshi2016,yu2016} by sweeping the relative detuning between the laser and cavity resonance from the blue- to the red-detuned \cite{Lamont13,herr2014}. The dynamics of mode-locking have been studied using various approaches to control the effective detuning, including laser frequency tuning \cite{herr2014,wang2016,li2017}, ‘power kick’ \cite{brasch2016,yi2015,webb2016}, and resonance frequency tuning using integrated heaters \cite{joshi2016} or free-carrier lifetime control \cite{yu2016}.

Recently, there has been interest in studying the nonlinear dynamics of bidirectionally pumped microresonators \cite{delbino2017,yang2017}. For the case in which the pumps have unequal powers, the counter-rotating fields experience different nonlinear phase shifts that leads to unequal detuning from the cavity resonances for the clockwise (CW) and counter-clockwise (CCW) directions. Such behavior can lead to bistability \cite{delbino2017} and can be exploited to create a gyroscope with enhanced sensitivity to rotation \cite{silver2017,wang2014}. For the case in which such a system can be mode-locked it would result in the generation of two soliton trains with different repetition rates in a single microresonator and thus be used as a dual-comb source in a number of applications \cite{dutt2016,yudual2016,suh2016,pavlov2017,bernhardt2010,coddington2016,link2017,millot2016}. Recently counter-propagating solitons were generated in silica microresonators using a single laser, frequency shifted using two acousto-optic modulators (AOM's) pumping a single microresonator \cite{yang2017}. The difference in effective detuning was controlled using the two AOM's and leads to a difference in repetition rate for the solitons. While there have also been recent demonstrations of bidirectional mode-locked solid state \cite{ideguchi16} and fiber \cite{kieu08,gowda2015} laser cavities, a microresonator-based system could be highly compact and fully integrated onto a chip. 


In this Letter, we present a novel approach to generating counter-rotating trains of solitons in a single microresonator using a single pump laser without using frequency shifting devices by thermally tuning the microresonator. By tuning the relative pump powers in the two directions, we can control the repetition rate of the two soliton-modelocked pulse trains. Such a dual comb source using a single pump laser and single microresonator eliminates common mode noise due to relative fluctuations between two resonators and lasers and would enable improved real-time, high signal-to-noise ratio (SNR) measurements of molecular spectra \cite{coddington2016}, time-resolved measurements of fast chemical processes \cite{fleisher2014}, and precise distance measurements \cite{coddington2009,suh2017}.

\begin{figure}[tb]
\centering
\includegraphics[width=\linewidth]{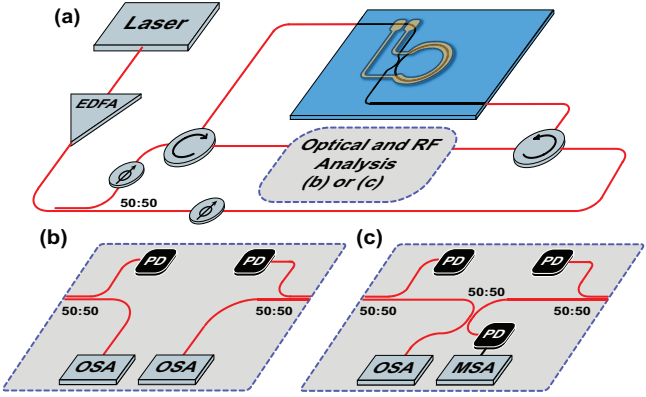}
\caption{(a) Experimental setup to generate counter-rotating solitons in a single microresonator using a single pump laser. We characterize the generated counter-rotating solitons (b) individually, measuring the optical spectra and transmitted optical powers in CW and CCW directions and (c) after combining the output in both directions to measure the mixed optical and heterodyned RF signal.}
\label{fig:setup}
\end{figure}

In our experiment (Fig.~\ref{fig:setup}(a)), we use a single-frequency laser (1559.79~nm) with a narrow linewidth (1~kHz) as our pump source, which is amplified using a polarization maintaining (PM) erbium-doped fiber amplifier (EDFA). The remaining experimental setup consists of PM components to ensure that the polarizations of the pump light and that of the generated combs are maintained throughout. We split the amplified output using a 50:50 splitter, and the outputs of the splitter are sent to a pair of variable optical attenuators (VOA's) to independently control the pump power in the CW and CCW directions. The pump for the CW and CCW directions is connected to port 1 of the two circulators. Port 2 of both circulators is connected to a pair of PM-lensed fibers to couple light in and out of the chip. We use a Si$_3$N$_4$ ring with a 200-GHz free spectral range (FSR) and a cross section of 950 x 1500 nm that yields anomalous group-velocity dispersion at the pump wavelength as required for soliton formation \cite{joshi2016}. The microresonator is undercoupled with an extinction ratio of 0.53, and the resonance frequency of the ring is controlled using integrated platinum resistive heaters. We observe a narrowing of the detuning region corresponding to simultaneous soliton generation in both directions that we believe occurs due to the nonlinear coupling between the counter-rotating modes. We find that the overlap in the detuning region for generation of the three-soliton state in both directions is sufficiently broad to permit stable operation, in contrast to the single- and two-soliton states where the detuning region was too narrow for sustained operation. The resulting 3-FSR comb spectra indicate three equally-spaced solitons in the cavity for each direction. The generated combs in the CW and CCW direction are coupled out using the lensed fibers at port 3 on the respective circulators, and the optical spectra and transmitted power of each are measured using two optical spectrum analyzers (OSA's) and fast photodiodes ($\geq$ 12.5~GHz) (Fig.~\ref{fig:setup}(b)). The two soliton trains are then combined using another 50:50 splitter, and the optical and RF properties of the dual comb are measured using an OSA and microwave spectrum analyzer (MSA) (Fig.~\ref{fig:setup}(c)). 

\begin{figure}[tb]
\centering
\includegraphics[width=\linewidth]{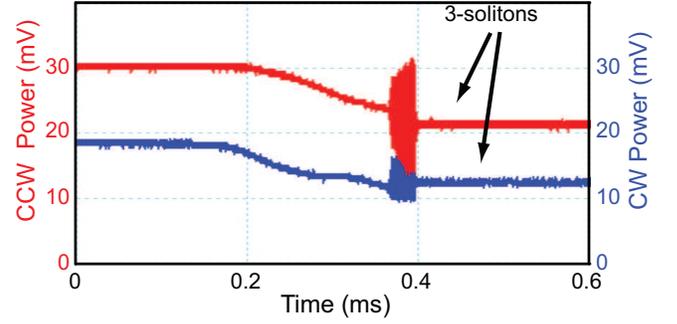}
\caption{The transmitted pump power as a downward current ramp is applied followed by a fixed current offset to reach the 3-soliton state. 30~mV corresponds to 3.86~mW in the bus waveguide.}
\label{fig:scanburst}
\end{figure}

We generate a 3-soliton state in both directions with resonance tuning of the microresonator at a speed of 200~Hz. In order to tune the cavity resonance frequency close to the pump laser frequency, we apply 98~mW of electrical power to the integrated heater (R~=~200~$\Omega$). The pump transmission is recorded as we scan the cavity resonance across the laser, and we observe a low-noise `step' on the red-detuned side characteristic of soliton mode-locking \cite{herr2014,joshi2016}, which corresponds to the 3-soliton state. We use the thermal tuning method to reach this state deterministically  \cite{joshi2016} by applying a downward current ramp to a fixed DC offset current (Fig.~\ref{fig:scanburst}). We generate a bidirectional 3-soliton state over pump powers from 1.35 to 6.1 mW in the bus waveguide in each direction and record its properties over this range. 

\begin{figure}[htb]
\centering
\includegraphics[width=\linewidth]{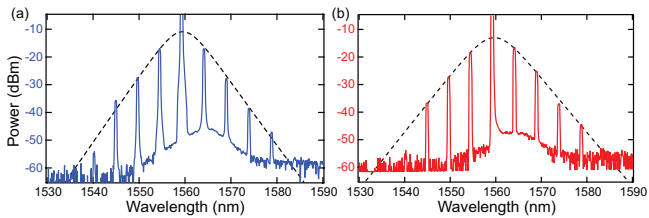}
\caption{Measured optical spectra for the (a) CW and (b) CCW directions. We observe a 3-soliton mode-locked comb in both directions, and the sech$^2$ fits are shown with the dashed black curves.}
\label{fig:specind}
\end{figure}

The generated combs are sent to a pair of OSA's to record the optical spectrum. To allow for simultaneous measurement of the spectra (Fig.~\ref{fig:specind}) and pump transmission (Fig.~\ref{fig:scanburst}), the OSA's are triggered using a signal from the arbitrary waveform generator that is used to drive the integrated heater. The 3-FSR spaced optical spectra in the CW (Fig.~\ref{fig:specind}(a)) and CCW (Fig.~\ref{fig:specind}(b)) directions show good agreement with the hyperbolic secant pulse profile, as shown by the black dashed curves.

\begin{figure}[tb]
\centering
\includegraphics[width=\linewidth]{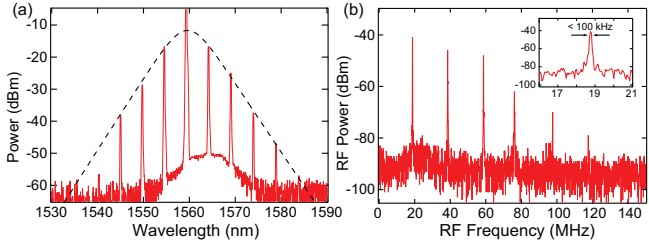}
\caption{(a) Optical spectrum for the dual comb with both the CW and CCW soliton trains combined. We observe a 3-soliton spectrum. The sech$^2$ fit is shown in the dashed black curve. (b) Measured heterodyned RF comb with a sequence of beat notes corresponding to multiples of 3$\times\Delta f_{\text{r}}$ for a power ratio $r$~=~0.67. The inset shows the first beat note in the heterodyned RF comb.}
\label{fig:mixedsoliton}
\end{figure}

We use a 50:50 splitter to combine the two combs and send them to both an OSA and a photodiode (bandwidth~$\geq$~250~MHz) to detect the heterodyne RF signal on a MSA. The measured optical spectrum (Fig.~\ref{fig:mixedsoliton} (a)) shows a hyperbolic secant spectral profile with a 3-FSR spacing as seen in the individual optical spectra for the each direction (Fig.~\ref{fig:specind}). However,  due to the OSA resolution limit of 1.25~GHz (0.01~nm), the difference in repetition rates is not resolvable. We measure the heterodyned RF signal and observe a RF comb with a spacing of 19~MHz, which indicates a difference in the FSR of 6.3~MHz since the measured RF beatnotes correspond to multiples of 3$\times \Delta f_{\text{r}}$ from the two 3-soliton states. The linewidth of the first RF comb line is $\leq$~100~kHz measured at a resolution bandwidth (RBW) of 50~kHz [inset of Fig. \ref{fig:mixedsoliton}(b)], which corresponds to a mutual coherence time for the two solitons of $\geq$~10~$\mu$s. 


We develop a simple model to predict the difference in repetition rates for the counter-rotating solitons. The self-phase modulation (SPM) and cross-phase modulation (XPM) for the pumps induces a shift in the pump-cavity detuning for the CW and CCW modes due to a change in the effective index. The pump-cavity detuning in each direction ($\delta \omega_{\text{CW}}$, $\delta \omega_{\text{CCW}}$) depends on pump detuning with respect to the cold cavity mode ($\delta \omega_{\text{p}}$) and the intracavity pump powers ($P_{\text{CW}},P_{\text{CCW}}$) as given by, 

\begin{equation}
\begin{aligned} \label{detuning}
\delta \omega_{\text{CW}} &= \delta \omega_{\text{p}} + \frac{\omega_0 n_2}{n_{\text{eff}} A_{\text{eff}}}(P_{\text{CW}} + 2 P_{\text{CCW}}), \\
\delta \omega_{\text{CCW}} &= \delta \omega_{\text{p}} + \frac{\omega_0 n_2}{n_{\text{eff}} A_{\text{eff}}}(P_{\text{CCW}} + 2 P_{\text{CW}}),
\end{aligned}
\end{equation}

\noindent where $\omega_0$ is the resonance frequency, $A_{\text{eff}}$ is the mode area, $n_{\text{eff}}$ is the effective index of the waveguide, and $n_2$ is the nonlinear index coefficient \cite{delbino2017}. Unequal pump powers in the CW and CCW modes yield a difference in the pump-cavity detuning in the CW and CCW direction. The peak power of the generated solitons has a linear dependence on the pump-cavity detuning \cite{herr2014,yi2015} as given by, \begin{align} \label{soliton}
P_{\text{sol}} = \frac{2 \ c \ A_{\text{eff}} \ \tau_{\text{r}}}{\omega_0 \ n_2 \ L}  \delta \omega
\end{align}

\noindent where $L$ is the cavity length, and $\tau_{\text{r}}$ is the round trip time. As a result, unequal pump powers leads to unequal peak powers for the counter-rotating solitons. We assume over each round trip the solitons acquire a nonlinear self-phase shift, as well as a cross-phase shift from the pump fields. Due to the small temporal overlap between the counter-rotating solitons we neglect the XPM from the counter-rotating soliton. If we assume the XPM from the pump fields acts on both solitons equally and cancels out, the unequal peak powers of the solitons in the two directions result in a difference in the nonlinear phase over one round trip that results in a difference in the repetition rates $\Delta f_{\text{r}}$ as given by, 

\begin{align} \label{delfsr}
\Delta f_{\text{r}} = |f_{\text{CW}} - f_{\text{CCW}}| = g \frac{n_2 \ f_{\text{r}}}{A_{\text{eff}} \ n_{\text{eff}}} |P_{\text{sol,CW}} - P_{\text{sol,CCW}}|,
\end{align} 
\noindent where $g$ is the factor for the nonlinear phase shift induced by the dissipative soliton on itself. 

Using Eqs. \ref{detuning}-\ref{delfsr}, $\Delta f_{\text{r}}$ can be expressed in terms of the transmitted pump power $P_{\text{out}}$ in the clockwise direction, the ratio $r = P_{\text{CCW}}/P_{\text{CW}}$ of pump powers, and the ratio $\eta$ of the intracavity pump power to the transmitted pump power, which depends on the losses in the ring and the coupling constant, as well as losses due to coupling from the bus waveguide to the lensed fiber (2 dB), at the circulator (1 dB) and 50:50 splitter (3 dB). The value of $\Delta f_{\text{r}}/P_{\text{out}}$ can then be expressed purely in terms of material and waveguide parameters such that,

\begin{align} \label{delfsrnorm}
\frac{\Delta f_{\text{r}}}{P_{\text{out}}} = g \frac{2 \ n_2 \ f_{\text{r}}}{n_{\text{eff}}\ A_{\text{eff}}} \ \eta \ |1-r|.
\end{align}

\begin{figure}[htb]
\centering
\includegraphics[width=\linewidth]{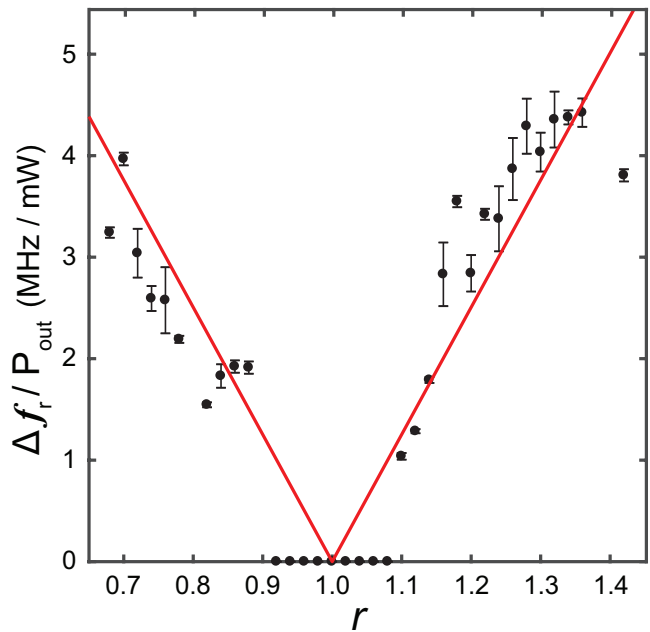}
\caption{The difference in repetition rate normalized to the power in the clockwise mode ($\Delta f_{\text{r}}/P_{\text{out}}$) as a function of the power ratio $r$. Each of the measured points from the experiment is plotted as a black dot. The red curve represents the theoretical curve from Eq. \ref{delfsrnorm}.}
\label{fig:rfbyp}
\end{figure}

This result suggests that we can control $\Delta f_{\text{r}}$ by simply varying the ratio of the counter-rotating pump powers. Experimentally we use the VOA's to independently control the pump power in the two directions. The coupled pump power in the bus waveguide in both directions is varied over a range of 1.35 to 6.1 mW. We measure the transmitted pump powers in each direction to determine $P_{\text{out}}$ and $r$. We measure the frequencies of the heterodyned RF peaks and infer the difference $\Delta f_{\text{r}}$. The ratio $\Delta f_{\text{r}}/P_{\text{out}}$ yields a normalized measure of the tunability of the difference in FSR at different power levels. We plot the measured values of $\Delta f_{\text{r}}/P_{\text{out}}$ and the fit to Eq. \ref{delfsrnorm} while varying $r$ in (Fig.~\ref{fig:rfbyp}). We observe reasonable agreement between the theoretically predicted curve and the measured values. The parameter $\eta$ in Eq. \ref{delfsrnorm} depends on the coupling constant between the bus waveguide and microresonator and on the waveguide loss. Over a range of power ratios close to unity, we observe locking between the two soliton trains, which is indicative of identical repetition frequencies for the CW and CCW soliton trains. A full understanding of the locking mechanism of the repetition rates over a range of power ratios will require extension of the theoretical analysis to include soliton comb formation dynamics including the coupling between the modes in both directions. We use Eq. \ref{delfsrnorm} to fit the red curve in Fig. \ref{fig:rfbyp} and from this fit extract the value of $\text{g} \times \eta$ to be 4600.  

In conclusion, we observe counter-rotating solitons in a single microresonator using a single pump laser. We demonstrate the ability to tune the difference in the repetition frequency of the two soliton trains by varying pump power for the modes in the clockwise and counterclockwise directions. Using a single-frequency laser and a single microresonator eliminates common mode noise in the dual-comb source. With future advances, we envisage a fully integrated tunable dual-comb source with electrical control of both the mode-locking as well as the tuning of repetition rates that would find applications in dual-comb spectroscopy and adaptive distance measurement.
\newline \newline
\noindent \textbf{Funding.} Air Force Office of Scientific Research (AFOSR) (FA9550-15-1-0303); National Science Foundation (NSF) (ECS-0335765); Defense Advanced Research Projects Agency (W31P4Q-15-1-0015); A.K. acknowledges a postdoc fellowship from the Swiss National Science Foundation (P2EZP2\_162288)
\newline \newline
\noindent \textbf{Acknowledgements.} This work was performed in part at the Cornell Nano-Scale Facility, a member of the National Nanotechnology Infrastructure Network, which is supported by the NSF.
  
\bibliography{counterprop}
\end{document}